\documentclass[iop,apj,appendixfloats]{emulateapj}
\usepackage{apjfonts}
\usepackage{graphicx}
\usepackage{amsmath}
\usepackage{amssymb}
\usepackage{ulem}

\newcommand{\se}[1]{\S\ref{sec:#1}}
\newcommand{\fig}[1]{Fig.~\ref{fig:#1}}

\newcommand{\be}{\begin{equation}}
\newcommand{\ee}{\end{equation}}
\newcommand{\bea}{\begin{eqnarray}}
\newcommand{\eea}{\end{eqnarray}}

\newcommand{\msun}{{\rm M}_\odot}
\newcommand{\Msun}{M_\odot}

\newcommand{\ifm}[1]{\relax\ifmmode#1\else$\mathsurround=0pt #1$\fi}
\newcommand{\kms}{\ifmmode\,{\rm km}\,{\rm s}^{-1}\else km$\,$s$^{-1}$\fi}

\newcommand{\Mpc}{\,{\rm Mpc}}
\newcommand{\kpc}{\,{\rm kpc}}
\newcommand{\pc}{\,{\rm pc}}
\newcommand{\Gyr}{\,{\rm Gyr}}

\newcommand{\K}{\,{\rm K}}
\newcommand{\cmc}{\,{\rm cm^{-3}}}
\newcommand{\cms}{\,{\rm cm^{-2}}}

\newcommand{\ltsima}{$\; \buildrel < \over \sim \;$}
\newcommand{\lsim}{\lower.5ex\hbox{\ltsima}}
\newcommand{\gtsima}{$\; \buildrel > \over \sim \;$}
\newcommand{\gsim}{\lower.5ex\hbox{\gtsima}}

\def\la{\langle}

\def\cmc{\,{\rm cm}^{-3}}
\def\cms{\,{\rm cm}^{-2}}

\def\M*{M_{\rm *}}
\def\Mv{M_{\rm v}}
\def\Rv{R_{\rm v}}

\def\cs{c_{\rm s}}

\def\NHI{N_{\rm HI}}

\newcommand{\angstrom}{\mbox{\normalfont\AA}}

\usepackage{color}
\lefthead{Mandelker et al.}
\righthead{Shattering of Cosmic Sheets}
\slugcomment{Submitted to ApJL}

\begin{document}
\vspace{1mm}

%--------------------------------------------------------
\title{Shattering of Cosmic Sheets due to Thermal Instabilities: a Formation Channel for Metal-Free Lyman Limit Systems}

% My first suggestion:

%Shattered Pancakes: Thermal Instabilities in Intergalactic Sheets Result in Metal-Free Lyman Limit Systems

% FB: Here are some alternative titles to consider:

% FB: Metal-Free Lyman Limit Systems resulting from the Shattering of Cosmic Sheets driven by Thermal Instabilities

% FB: Shattering of Cosmic Sheets due to Thermal Instabilities: a Formation Channel for Metal-Free Lyman Limit Systems

% FB: Shattering Action between the Sheets: Thermal Instabilities and the Formation of Metal-Free Lyman Limit Systems

%--------------------------------------------------------

\author{Nir Mandelker\altaffilmark{1,2,3}, 
Frank C. van den Bosch\altaffilmark{2},
Volker Springel\altaffilmark{4,3},
Freeke van de Voort\altaffilmark{4,2,3},
\vspace{8pt}}

\altaffiltext{1}
{corresponding author: nir.mandelker@yale.edu}
\altaffiltext{2}
{Department of Astronomy, Yale University, PO Box 208101, New Haven, CT, USA}
\altaffiltext{3}
{Heidelberger Institut f{\"u}r Theoretische Studien, Schloss-Wolfsbrunnenweg 35, 69118 Heidelberg, Germany}
\altaffiltext{4}
{Max Planck Institute for Astrophysics, Karl-Schwarzschild-Str{\ss}e 1, D-85748 Garching, Germany}

\begin{abstract}
We present a new cosmological zoom-in simulation, where the zoom region consists of two halos with virial mass $\Mv\sim 5\times 10^{12}\msun$ and a $\sim \Mpc$ long cosmic filament connecting them at $z\sim 2$. Using this simulation, we study the evolution of the intergalactic medium in between these two halos at unprecedented resolution. At $5\gsim z \gsim 3$, the two halos are found to lie in a large intergalactic sheet, or ``pancake'', consisting of multiple co-planar dense filaments along which nearly all halos with $\Mv>10^9\msun$ are located. 
%These filaments merge within the sheet, forming the single large filament connecting the two halos at $z\gsim 2$. 
This sheet collapses at $z\sim 5$ from the merger of two smaller sheets. The strong shock generated by this merger leads to thermal instabilities in the post-shock region, and to a shattering of the sheet resulting in $\lsim \kpc$ scale clouds with temperatures of $T\gsim 2\times 10^4\K$ and densities of $n\gsim 10^{-3}\cmc$, which are pressure confined in a hot medium with $T\sim 10^6\K$ and $n\gsim 10^{-5}\cmc$. When the sheet is viewed face on, these cold clouds have neutral hydrogen column densities of $\NHI>10^{17.2}\cms$, making them detectable as Lyman limit systems, though they lie well outside the virial radius of any halo and even well outside the dense filaments. Their chemical composition is pristine, having zero metalicity, similar to several recently observed systems. Since these systems form far from any galaxies, these results are robust to galaxy formation physics, resulting purely from the collapse of large scale structure and radiative cooling, provided sufficient spatial resolution is available.
\end{abstract} 
 

%%%%%%%%%%%%%%%%%%%%%%%%%%%%% 
\section{Introduction} 
\label{sec:intro}

\smallskip
%
%There have recently been several discoveries of Lyman limit systems (LLSs), gas clouds with neutral Hydrogen column densities $10^{20.3}>\NHI>10^{17.2}$ making them optically thick bluewards of the Lyman limit, $\lambda_{\rm LL}\lsim$ 912\angstrom,
%
% In the standard $\Lambda$CDM cosmological model, galaxies form within the cosmic web, a network of filaments and sheets that comprise the distribution of matter on the largest scales. 
%
% Observational and theoretical studies of the circumgalactic medium (CGM) and the intergalactic medium (IGM), the gas surrounding galaxies within and outside of their dark matter halos, have witnessed remarkable advances in recent years (see \citealp{Tumlinson17} for a recent review). In addition to being a major reservoir for baryons, the CGM and IGM play key roles in galaxy evolution and star-formation, through cycles of gas accretion, outflows, and recycling. 
%
% One of the primary goals of modern astronomy is to map the distribution of baryons in the Universe. 
%
Only a small fraction of the Universe's baryons and metals belong to galaxies \citep[e.g.][]{Tumlinson17,Wechsler18}. The rest %are thought to 
reside in the circumgalactic medium (CGM), the space outside galaxies but %still 
within their host dark matter halo, and the intergalactic medium (IGM), the space in between dark matter halos. Both of these baryonic reservoirs are intimately linked to galaxy evolution through cycles of gas accretion, star-formation, galactic outflows, and eventual re-accretion \citep[e.g.][]{Putman12,McQuinn16,Tumlinson17}. Thus, the physical properties and chemical composition of the IGM and CGM %gas 
offer valuable insight into processes related to galaxy formation and evolution. 

\smallskip
In recent decades, the low density gas in the IGM and CGM has been probed using absorption line spectroscopy along lines of sight to distant QSOs or galaxies \citep[e.g.][]{Lynds71,Hennawi06,Steidel10}. 
%This technique allows one to infer the neutral gas content and metal content of intervening gas clouds.
Intervening gas clouds with low neutral hydrogen column densities, $\NHI\lsim 10^{15}\cms$, are understood to reside in the IGM and comprise the Lyman-$\alpha$ Forest, hereafter LyAF. %, \citep{McQuinn16}. 
This gas is thought to trace fluctuations in the underlying dark matter distribution which are still in the linear regime, making diagnostics of the LyAF a powerful tool to constrain cosmology (see \citealp{Rauch98} and \citealp{McQuinn16} for reviews). Clouds with high column densities, $\NHI> 10^{17.2}\cms$, are optically thick bluewards of the Lyman limit, $\lambda<$ 912\angstrom, and are referred to as Lyman Limit Systems %\footnote{Formally, LLSs are clouds with $10^{17.2}<\NHI< 10^{20.3}\cms$, while higher column density clouds are known as damped Lyman-$\alpha$ absorbers (DLAs).} 
(LLSs). At redshifts $2\lsim z\lsim 5$, LLSs exhibit a broad distribution of metalicities. The bulk of the population has $Z\sim 10^{-2}\,Z_{\odot}$, while a handful of systems have $Z< 10^{-3}\,Z_{\odot}$ \citep{Fumagalli16,Lehner16,Robert19}.

\smallskip
LLSs, particularly those with $Z> 10^{-3}\,Z_{\odot}$, are commonly thought to reside in the CGM %of galaxies
rather than the IGM \citep{Sargent89,Fumagalli16,Lehner16}. However, the recent discovery of several LLSs with $Z<10^{-3.4}\,Z_{\odot}$ as well as potentially pristine LLSs at $3<z<5$ has led some to question whether these may represent a separate population originating in the IGM \citep{Fumagalli11a,Crighton16,Robert19}. %Pollution from even 
A single PopIII supernova would pollute gas to higher metalicity values \citep{Wise12,Crighton16}, and simulations of structure formation which include PopIII star formation suggest that such low metalicities %can 
exist only in isolated low-density patches of the IGM \citep[e.g.][]{Tornatore07,Wise12}. %Another possibility is that pristine LLSs originate in cold
Alternatively, pristine LLSs may originate in cold accretion streams feeding massive galaxies from the IGM \citep{Dekel09,Fumagalli11b,FG11b,vdv12}. While cosmological simulations suggest that the typical metalicity in %streams within the CGM of massive galaxies at $z<5$ is $\gsim 10^{-3}\,Z_{\odot}$
such streams is $\gsim 10^{-3}\,Z_{\odot}$ at $z<5$ \citep{vdv12b,Ceverino15c,M18}, lower metalicity clouds may still be present. However, it has also been suggested that the evolution of the number of LLSs per unit redshift at $z>3.5$ is inconsistent with a contribution from the CGM alone, indicating a growing contribution of LLSs in the IGM above this redshift \citep{Fumagalli13}. All in all, the origin of extremely metal-poor Lyman limit systems in the IGM is not yet understood.

\smallskip
It is notoriously difficult to study the detailed properties of gas in the IGM and CGM with cosmological simulations. The resolution in most state-of-the-art simulations is adaptive in a quasi-Lagrangian sense, such that the effective mass resolution is fixed. %As a consequence, 
Consequently, the spatial resolution becomes very poor in the low density CGM and even worse in the IGM \citep{Nelson16}, orders of magnitude larger than the cooling length of $T\sim 10^4\K$ gas, $l_{\rm cool}=\cs t_{\rm cool}\sim 100\pc~(n/10^{-3}\cmc)^{-1}$, where $\cs$ is the sound speed and $t_{\rm cool}$ is the cooling time (\citealp{McCourt18}, hereafter M18; \citealp{Sparre19}). While several groups have recently introduced different methods to better resolve the CGM \citep{Hummels18,Corlies18,Peeples19,Suresh19,vdv19}, we are unaware of similar attempts to better resolve the IGM. %improve the resolution in the IGM. 
% Mention standard zooms where zoom-in region is limited to <~2*Rvir around galaxy?

\smallskip
In this letter, we introduce a new cosmological simulation where we zoom-in on a large region of the IGM in between two massive galaxies at $z\sim 2.3$, with a comoving separation of $\sim 3\Mpc/h$. This is the highest resolution simulation of such a large patch of the IGM to date. Using this simulation, we investigate the evolution of the IGM and show how thermal instabilities triggered by shocks during the collapse of large-scale structure can lead to the formation of pristine LLSs, far from any galaxies. The simulation is described in \se{sim}. In \se{results} we present our results, and we conclude in \se{conc}. Throughout, we assume a flat $\Lambda$CDM cosmology with $\Omega_{\rm m}=1-\Omega_{\Lambda}=0.3089$, $\Omega_{\rm b}=0.0486$, $h=0.6774$, $\sigma_8=0.8159$, and $n_{\rm s}=0.9667$ \citep{Planck16}.

%%%%%%%%%%%%%%%%%%%%%%%%%%%%% 
\section{Simulation Method} 
\label{sec:sim}

\smallskip
We perform simulations using the quasi-Lagrangian moving-mesh code \texttt{AREPO} \citep{Springel10}. To select our target halos, we first consider the 200 most massive halos at $z\sim 2.3$ in the Illustris TNG100\footnote{http://www.tng-project.org} magnetohydrodynamic cosmological simulation \citep{Pillepich18b,Nelson18,Springel18}. These span a mass range of $\Mv\sim (0.7-27)\times 10^{12}\msun/h$, where $\Mv$ is the virial mass defined using the \citet{Bryan98} spherical overdensity. We then select all pairwise combinations of them with a comoving distance in the range $(2.5-4.0)\Mpc/h$, finding 48 such halo pairs. Visual inspection revealed each such halo pair to be directly connected by a dark matter cosmic web filament, with comparable radius to the halo virial radii. %of these to be connected by a large %intergalactic filament along the cosmic web.
%cosmic web filament. 
One such pair was randomly chosen for resimulation, consisting of two halos with $\Mv\sim 3.4\times 10^{12}\msun/h$ each, separated by a proper distance of $D\sim 0.8\Mpc/h$. At $z=0$, the two halos have masses $\Mv\sim (1.1-1.3)\times 10^{13}\msun/h$ and are $\sim 1.8\Mpc/h$ apart, so their comoving distance has decreased by $\lsim 30\%$.

\smallskip
%We define $R_{\rm ref}=1.5\times{\rm max}(\Rv)\sim 163\kpc/h$, 1.5 times the larger of the two virial radii %where the maximum is between the two halos 
We define $R_{\rm ref}=1.5\times R_{\rm v,max}\sim 163\kpc/h$, with $R_{\rm v,max}$ the larger of the two virial radii at $z=2.3$. The zoom-in region is %defined as 
the union of a cylinder with radius $R_{\rm ref}$ and length $D$ extending between the two halo centers, and two spheres of radius $R_{\rm ref}$ centred on either halo. We trace all dark matter particles within this volume back to the initial conditions of the simulation, at $z=127$, refine the corresponding Lagrangian region to higher resolution, and rerun the simulation to $z=2$, when the region of interest by construction becomes contaminated by low resolution material from outside the refinement region. The simulations were performed with the same physics model used in the TNG100 simulation, described in detail in \citet{Weinberger17} and \citet{Pillepich18}. We briefly summarize below the implementation of the ionizing radiation field and of cooling, which are most relevant to our current work. 

\smallskip
We follow the production and evolution of nine elements (H, He, C, N, O, Ne, Mg, Si, and Fe). These are produced in supernovae Type Ia and Type II and in AGB stars according to tabulated mass and metal yields. Metal line cooling is included using pre-calculated values as a function of density, temperature, metalicity and redshift, with corrections for self-shielding  \citep{Wiersma09}. The metal enriched gas radiatively cools in the presence of a spatially uniform but redshift dependent ionizing UV background \citep[UVB;][]{FG09}, which is instantaneously switched on at $z=6$. To minimize any potential influence of this instantaneous switching on of the UVB, we limit our current analysis to $z\le 5$. Cooling is further modulated by the radiation field of nearby active galactic nuclei (AGN) by superimposing the UVB with the AGN radiation field \citep{Vogelsberger13}.

\smallskip
We performed five simulations with different resolution within the refinement region. A detailed convergence study will be presented in an upcoming paper (Mandelker et al., in prep.). In the current letter we focus on our highest resolution simulation, which has a dark matter particle mass of $m_{\rm dm}=8.2\times 10^4\,\msun$ and a Plummer-equivalent gravitational softening of $\epsilon_{\rm dm}=250\pc$ comoving. Gas cells are refined such that their mass is within a factor of 2 of $m_{\rm gas}=1.5\times 10^4\msun$, and have a minimal gravitational softening $\epsilon_{\rm gas}=0.5\epsilon_{\rm dm}$. 
We compare this to a simulation with comparable resolution to TNG100, having $m_{\rm dm}=5.3\times10^6\msun$, $m_{\rm gas}=1.0\times10^6\msun$, and $\epsilon_{\rm dm}=2\epsilon_{\rm gas}=1000\pc$ comoving.
%, and $\epsilon_{\rm gas}=500\pc$ comoving.
%The corresponding values in the parent TNG100 simulation are $m_{\rm dm}=7.5\times10^6\msun$, $\epsilon_{\rm dm}=740\pc$ comoving, $m_{\rm gas}=1.4\times10^6\msun$, and $\epsilon_{\rm gas}=190\pc$ comoving. 

%\Fig{halos} shows the evolution of the virial masses, radii, and physical separation between the two main halos as a function of cosmic scale factor, $a=(1+z)^{-1}$. These results are independent of resolution.

%%\begin{figure*}
%%\begin{center}
%%\includegraphics[width =0.49 %\textwidth]{halo_masses.png}
%%\hspace{-0.3cm}
%%\includegraphics[width =0.49 %\textwidth]{halo_sizes.png}
%%\end{center}
%\begin{figure}
%\includegraphics[trim={0 0 0 1.0cm}, clip, width =0.49 \textwidth]{masses_and_sizes.png}
%\caption{Properties of the two main halos in the simulation as a function of the cosmic expansion factor $a=(1+z)^{-1}$. Solid (dashed) black and red lines show the halo virial masses (radii). Halo 1 is the more massive of the two at the final snapshot of the simulation, $z\sim 2.0$. At this time, both halos have virial masses $\Mv\sim 5\times 10^{12}\msun$. The dashed blue line shows the physical distance between the two halos, which is $D\sim 1.1\Mpc$ at $z=2$.}
%\label{fig:halos} 
%\end{figure}

%%%%%%%%%%%%%%%%%%%%%%%%%%%%% 
\section{Results} 
\label{sec:results}

\begin{figure*}
\begin{center}
\includegraphics[trim={1.45cm 0.4cm 8.5cm 0.0cm}, clip, width =0.98 \textwidth]{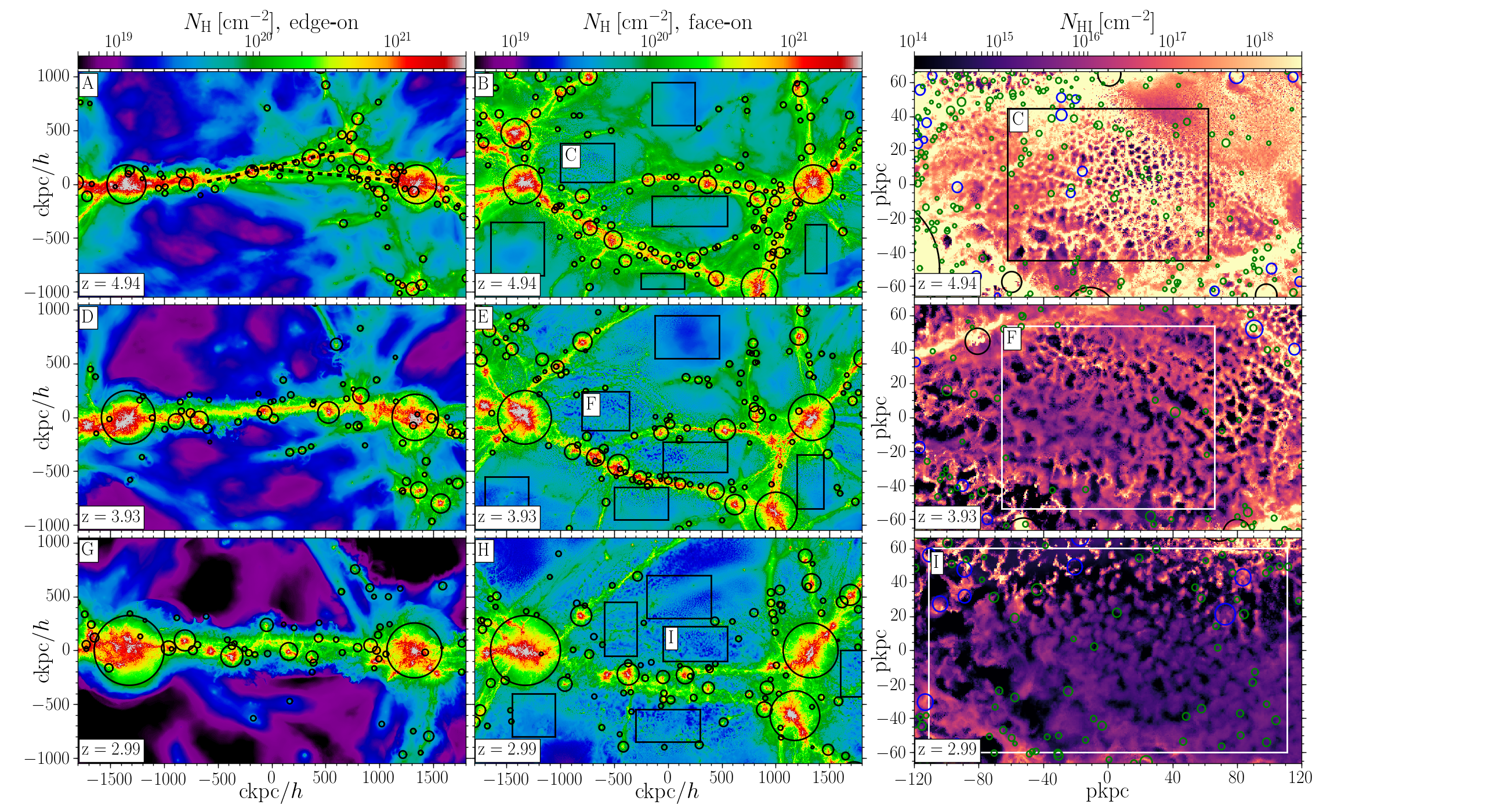}
\end{center}
\caption{The %zoom-in region of the simulation 
simulation zoom region at $z\sim 5$ (top), $z\sim 4$ (middle) and $z\sim 3$ (bottom). The left and center columns show the large scale structure %in between 
surrounding the two main halos, which lie in a cosmic sheet %with Halo1 on the right and Halo2 on the left, at $\sim \pm 1400~{\rm ckpc}/h$. 
shown edge-on and face-on in the left and center columns, respectively.
%The left panels show edge-on views of this sheet while the center panels show face-on projections. 
The color scale indicates the total hydrogen column density, $N_{\rm H}$, integrated over $\pm 400~{\rm ckpc}/h$. Black circles mark dark matter halos with virial mass $\Mv>10^9\msun$, while their sizes denote $\Rv$. Nearly all these halos lie along dense filaments within the sheet. 
The right column zooms in on the IGM regions %of the IGM 
marked with black rectangles in the centre column, and shows the neutral Hydrogen column density, $\NHI$, integrated over $\pm 30~{\rm pkpc}$ at $z\sim 5$ and $z\sim 4$, and $\pm 60~{\rm pkpc}$ at $z\sim 3$, roughly where the vertical gas density profile reaches the universal mean. Note the different spatial units in these panels. Black, blue, and green circles mark halos with $\Mv>10^9\msun$, $10^9>\Mv>10^8\msun$, and $10^8>\Mv>10^7\msun$ respectively, their sizes denoting $\Rv$. A collision between two initially inclined sheets at $z\sim 5$ (dashed lines in panel A), produces a strong shock which leads to thermal instability in the post-shock region within the merged sheet. This causes the sheet to ``shatter'' and produces a multiphase medium.}
\label{fig:NH}
\end{figure*}

\smallskip
In \fig{NH} we show the evolution of the large scale structure surrounding our system, at $z\sim 5$, $4$, and $3$. The left and center columns show the total hydrogen column density, $N_{\rm H}$, in two orthogonal projections, with the intergalactic sheet containing the two halos shown edge-on and face-on, respectively. At $z>5$ the system actually consists of two sheets initially inclined to one another, marked by dashed lines in panel A. %still visible on the right side of panel A. %The two sheets
These merge at $z\sim 5$, with only a single sheet visible in panels D and G. The sheet contains several prominent coplanar filaments, with end-points at either of the two main halos and along which lie nearly all halos with $\Mv>10^9\Msun$. %are located. 
Most of these filaments merge at $z<3$, leaving behind the single giant filament %which was 
selected at $z=2.3$. The beginning of this merger %can be seen 
is visible in panel H. 
We note that the configuration of our system at $z\sim 3$ is remarkably similar to a system recently observed at $z\sim 3.2$ with MUSE \citep{Lusso19}.
%We note that at $z\sim 3$, our simulated system of two massive halos connected by a $\lsim \Mpc$ scale filament with a third large halo nearby (bottom-right of panel H) is remarkably similar to a system recently observed at $z\sim 3.23$ with MUSE \citep{Lusso19}.

\begin{figure*}
\begin{center}
\includegraphics[trim={4.4cm 1.2cm 3.7cm 0.9cm}, clip, width =0.98 \textwidth]{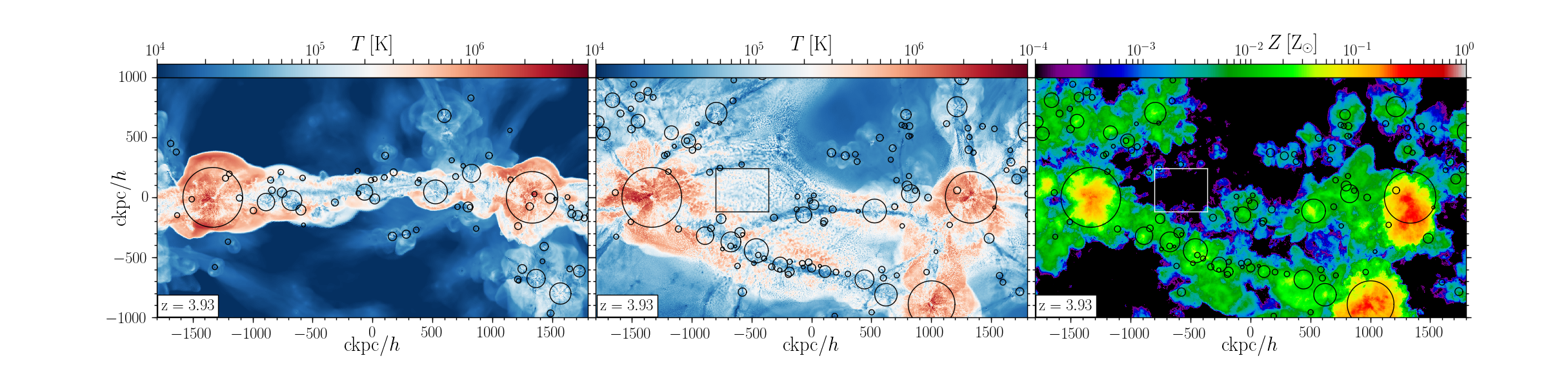}
\end{center}
\caption{Left and center: Temperature map of the system at $z=4$, in the same frame as panels D and E in \fig{NH}. We show the density-weighted average temperature along the line of sight, integrated over $\pm 400~{\rm ckpc}/h$. The accretion shock at the sheet edge is clearly visible in the edge-on view. In the face-on view, %we see that the filaments remain cold, 
the filaments appear cold while the sheet regions have a multiphase structure with hot and cold regions owing to thermal instability. Right: Metalicity map in the same frame as the center panel. The filaments are enriched to $Z\gsim 10^{-2}{\rm Z_{\odot}}$, while the sheet regions retain $Z<10^{-3}{\rm Z_{\odot}}$.
}
\label{fig:Temp}
\end{figure*}

\begin{figure*}
\begin{center}
\includegraphics[trim={0.10cm 1.5cm 1.85cm 1.85cm}, clip, width =0.98 \textwidth]{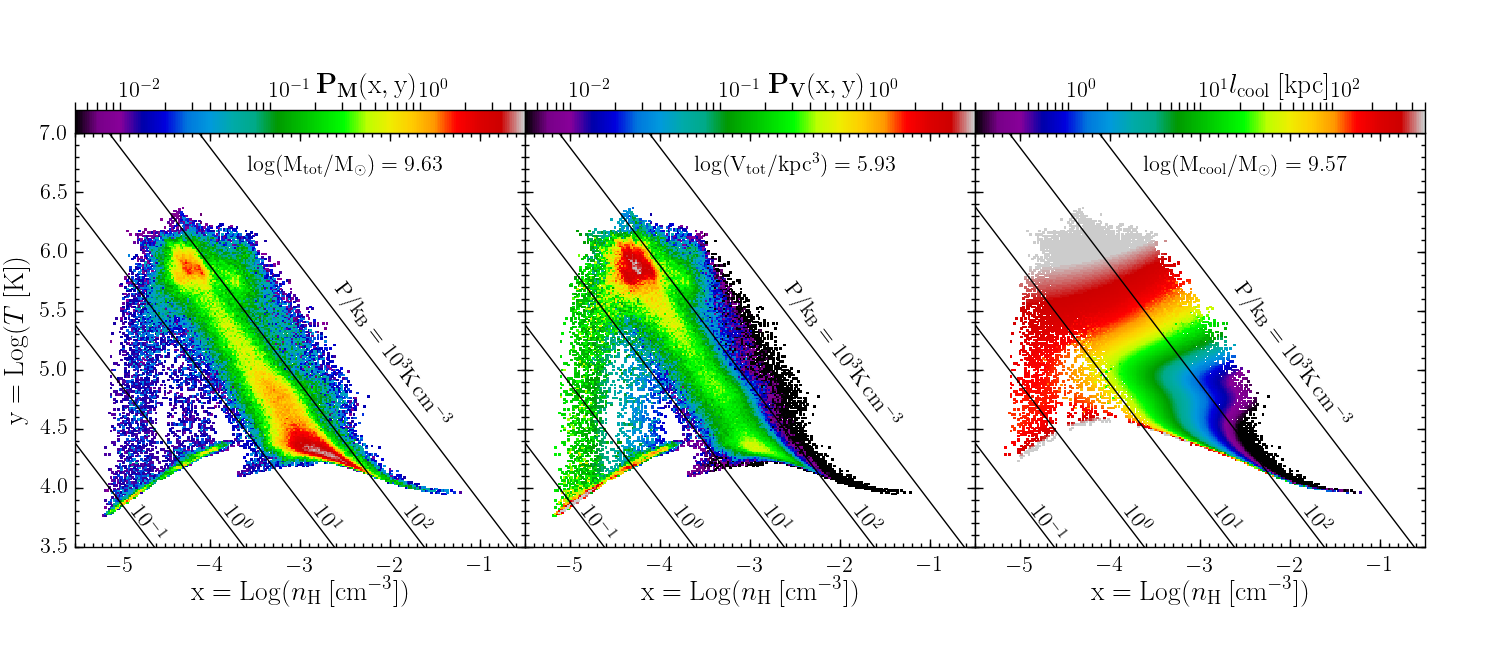}
\end{center}
\caption{Phase diagrams for the gas at $z=4$ in region F from \fig{NH}. The left and center panels show the mass- and volume-weighted distributions, respectively. The total mass and total volume are listed in each panel. The color represents a normalized probability density, such that the integral of ${\bf{P}}(x,y)\,{\rm d}x\,{\rm d}y$ over the full range of parameter space equals unity, where $x={\rm log}(n_{\rm H})$ and $y={\rm log}(T)$. Diagonal lines show constant thermal pressure, from $P_{\rm th}/k_{\rm B}=0.1-1000~{\rm K~cm^{-3}}$, as marked. The adiabat of the pre-shock gas is visible at low densities and low temperatures. Following the shock, the gas concentrates along an isobar with $P_{\rm th}\sim 50~\K~\cmc$. Two phases are clearly visible, with most of the mass/volume having $(n_{\rm H},T)\sim (10^{-3}\cmc,10^{4.5}\K)/(10^{-4.5}\cmc,10^6\K)$. %In the right panel we show 
The right panel shows the cooling length, $l_{\rm cool}=\cs t_{\rm cool}$, for gas in the same region, considering only gas undergoing net cooling. The total mass of the cooling gas is listed. The cooling length in the post-shock gas with $T\sim 2\times 10^5$ is $l_{\rm cool}\sim 10\kpc$, while in the cold phase it is $\sim 1\kpc$.}
\label{fig:phase_diagrams}
\end{figure*}

\begin{figure*}
\begin{center}
\includegraphics[trim={1.0cm 1.3cm 3.0cm 1.2cm}, clip, width =0.98 \textwidth]{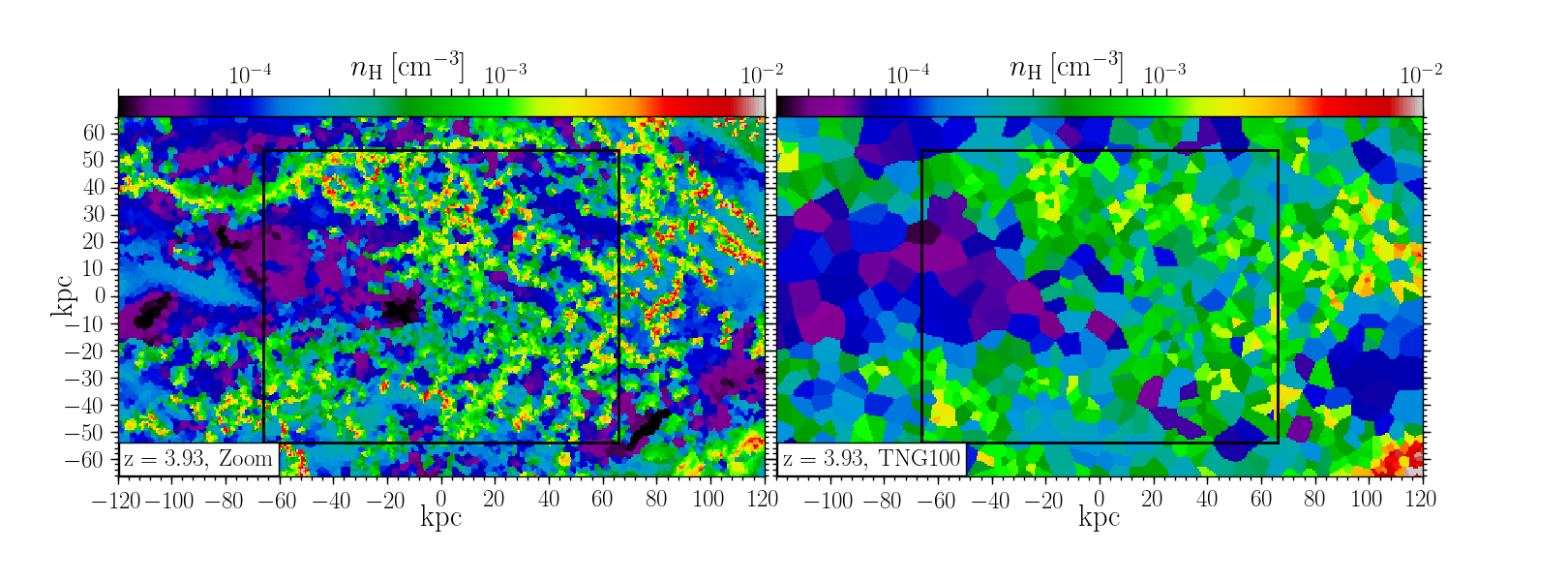}
\end{center}
\caption{Shattering of the sheet due to thermal instability. We show an infinitesimally thin slice of the gas density near the midplane of the sheet at $z=4$, in region F (as in \fig{phase_diagrams}). The left panel shows our highest resolution simulation, while the right panel shows a simulation with comparable resolution to TNG100. The typical cell size is $\sim 800\pc$ in the former and $\sim 4\kpc$ in the latter. The cooling length in the post-shock medium, with $T\gsim 10^5\K$ and $n_{\rm H}\lsim 10^{-3}\cmc$, is $l_{\rm cool}\sim 10\kpc$. This is well resolved in our high resolution simulation which allows the shattering process to begin, but not in TNG100 which is why the resulting density distribution is much smoother, and in particular no LLSs are produced. %However, the cooling length in the cold phase, with $T\sim 2\times10^4\K$ and $n_{\rm H}\lsim 10^{-2.5}\cmc$, is $l_{\rm cool}\sim 1\kpc$ which is only marginally resolved in our high resolution simulation, suggesting these results may not yet be converged.
}
\label{fig:slice}
\end{figure*}

\begin{figure}
\includegraphics[width =0.49 \textwidth]{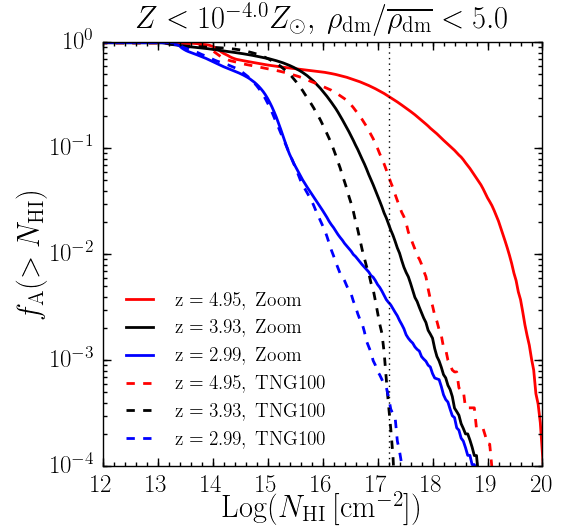}
\caption{
Covering fraction of pristine Ly$\alpha$ absorbers as a function of $N_{\rm HI}$ in regions C, F, and I from \fig{NH}, corresponding to $z\sim 5$, $4$, and $3$ and shown %as red, black, and blue lines respectively. 
in red, black, and blue respectively. Solid lines show our high-resolution simulation while dashed lines show a simulation with TNG100 resolution. The $y$-axis shows the fraction of surface area with %${\rm HI}$ column density 
$\NHI$ greater than the value on the $x$-axis due to pristine gas cells with $Z<10^{-4}\,Z_{\odot}$ in low density regions with $\rho_{\rm dm}<5{\overline{\rho_{\rm dm}}}$, %where ${\overline{\rho_{\rm dm}}}$ is the mean 
with ${\overline{\rho_{\rm dm}}}$ the mean matter density at the corresponding redshift. The vertical dotted line marks the LLS threshold. In our high resolution simulation, the covering fractions of pristine LLS are $\sim 0.005$, $0.01$, and $0.35$ at $z\sim 3$, $4$, and $5$, respectively. In the simulation with TNG100 resolution, the covering fractions are $\sim 4\times 10^{-4}$, $4\times 10^{-4}$, and $0.05$, respectively.
}
\label{fig:convergence} 
\end{figure}

\smallskip
Following the %merger of the two sheets 
sheet collision at $z\sim 5$, several regions between the filaments in the post-merger sheet develop a granular morphology. As the merging sheets were initially inclined, the collision and %the
resulting granular structure propagate from left-to-right in \fig{NH}, as can be seen by comparing panels B and E. In panels B, E, and H we highlight such granular regions, denoted C, F, and I, selected %not to contain any 
to contain no halos with $\Mv>10^9\msun$ and to not intersect any filaments. In the right-hand column %panels C, F, and I, 
we show the neutral hydrogen column density, $N_{\rm HI}$, within these regions. The granular structure is even more prominent in $N_{\rm HI}$, with many regions being classified as LLSs, %with 
$N_{\rm HI}>10^{17.2}\cms$. Importantly, these regions do not coincide with the locations of %any 
dark matter halos, or with %overdensities 
fluctuations in the dark matter distribution which is smooth in these regions. We examined several %additional
similar %granular 
regions within the sheet %not intersecting with any filaments or $\Mv>10^9\msun$ halos, 
in each snapshot, also marked in panels B, E, and H, and found the gas properties to be very similar in all regions within the post-merger sheet.% of them. 
%We hereafter focus only on the regions highlighted in \fig{NH}. 

\smallskip
As we argue below, this granular morphology seems to be triggered by nonlinear thermal instabilities within the post-shock sheet. Based on the model presented in M18, nonlinear thermal instabilities in a rapidly cooling medium cause the medium to ``shatter'', forming dense cloudlets with $T\gsim 10^4\K$ in pressure equilibrium with a more tenuous, hot background. The size of these cloudlets is set by the local cooling length, $l_{\rm cool}=\cs t_{\rm cool}$. This procedure is hierarchical, in the sense that as the gas cools, $l_{\rm cool}$ decreases, causing existing cloudlets to shatter into even smaller cloudlets. We note that there are several %qualitative 
differences between our system and the idealized study of M18. Firstly, the gas in our system is photo-heated by the UV background, while M18 considered a purely cooling system in collisional-ionization equilibrium. Second, in M18 the external pressure is set by the thermal pressure in the hot background, where the cooling time is assumed to be much longer than the shattering timescale in the cooling medium, while in our case it is set by the ram pressure of the infalling material, as discussed below. Finally, our system is in 3D while those studied in M18 were 2D. Nevertheless, as we argue below our results appear consistent with the %predictions of the 
M18 shattering model.

\smallskip
In \fig{Temp} we show the projected, density-weighted gas temperature in the same frame as panels D and E from \fig{NH}. A planar accretion shock around the sheet, triggered by the earlier collision, is clearly visible in the edge-on view, as are spherical accretion shocks around the two main halos. In the face-on view, the filaments appear cold, with $T\sim 2\times 10^4\K$, while the regions between filaments exhibit a multiphase structure, with hot and cold regions coexisting in a granular structure similar to that seen in the column density (\fig{NH}). %maps in \fig{NH}. 
In the right-hand panel of \fig{Temp}, we show the projected metalicity in the face-on view of the sheet at $z\sim 4$. While the filaments are enriched to $Z\gsim 10^{-2}{\rm Z}_{\rm \odot}$, consistent with previous studies \citep{vdv12b,Ceverino15c,M18}, the regions in between the filaments retain near-pristine compositions with $Z< 10^{-3}{\rm Z}_{\rm \odot}$, due to their large distance from any star-forming galaxies.

\smallskip
%In \fig{phase_diagrams} we show 
\fig{phase_diagrams} shows the distribution of gas within region F from \fig{NH} in density-temperature space, weighted by mass (left) and by volume (middle). The post-shock gas has roughly constant pressure, $P_{\rm th}\sim 50~\K~\cmc$, roughly the ram pressure of the infalling gas which has a density of $\sim 4\times 10^{-29}{\rm g~\cmc}$ and a velocity of %$(200-250)\kms$. 
$\sim 200\kms$. Two phases in approximate pressure equilibrium are apparent, with most of the mass at $(n_{\rm H},T) \sim (10^{-3}\cmc,10^{4.5}\K)$, and most of the volume at $\sim (10^{-4.5}\cmc,10^6\K)$. In the right panel we show the cooling length, $l_{\rm cool}=\cs t_{\rm cool}$, %as a function of gas density and temperature 
for gas undergoing net cooling. %ignoring gas which is undergoing net heating due to the UV background. 
The post-shock gas, which has a temperature of $T\sim 2\times 10^5\K$, has a cooling length of $l_{\rm cool}\sim 10\kpc$, while the cold phase along the same isobar has $l_{\rm cool}\sim 1\kpc$. 

\smallskip
In \fig{slice} we show the hydrogen density in a slice through the sheet-midplane %of the sheet region highlighted above, 
in region F from \fig{NH}. On the left we show %the results from 
our highest resolution simulation, while on the right we show %results from 
a simulation with similar resolution to TNG100. In the former, the typical (minimal) cell size within this region is $\Delta\sim 0.8~(0.3)\kpc$, %while the minimal cell size is $\sim 0.3\kpc$,
significantly smaller than the post-shock cooling length and comparable to the cooling length in the dense phase. We are thus able to resolve the onset of shattering %of the sheet 
into dense $\sim \kpc$ scale cloudlets (M18), resulting in the large neutral column densities seen in \fig{NH}. However, since the minimal cooling length is only marginally resolved, the end result is not converged and the actual cloudlets are expected to be smaller. On the other hand, in the simulation with TNG100 resolution, the typical (minimal) cell size within this region is $\Delta\sim 4.0~(2.5)\kpc$. %while the minimal cell size is $\sim 2.5\kpc$. 
The initial phases of the shattering are thus unresolved, and no dense cloudlets are formed. We note that the thermal Jeans length in the cold phase is $L_{\rm J}=[9c_{\rm s}^2/(4\pi G\rho)]^{1/2}\sim 30\kpc$, significantly larger than the cooling length, the cloud sizes, and the typical cell size. This implies that the clouds are not the result of gravitational instability in the sheet, and supports our hypothesis that they result from thermal instabilities and shattering.

\smallskip
In \fig{convergence} we show the covering fraction of neutral hydrogen as a function of $N_{\rm HI}$, in the two simulations shown in \fig{slice}. We show results at $z\sim 5$, $4$, and $3$, corresponding respectively to regions C, F, and I in \fig{NH}. In order to focus on gas with primordial composition that condenses due to thermal instabilities %in the post-shock sheet
rather than %due to the underlying dark matter or halo distribution, 
fluctuations in the underlying dark matter distribution as is often assumed in studies of the LyAF, 
when evaluating $N_{\rm HI}$ we ignore all cells with metalicity $Z>10^{-4}\,Z_{\odot}$ and with a dark matter density greater than 5 times the Universal mean at the relevant redshift. At all redshifts, the covering fraction of LLSs is significantly larger in our high resolution simulation than in the simulation with TNG100 resolution. As discussed above, this is because the latter does not resolve the initial shattering of the post-shock medium within the sheet. At $z\lsim 4$ we find covering fractions of order $\sim 1\%$ for pristine LLSs. At $z\sim 5$ the covering fraction is $\sim 35\%$. We find comparable covering fractions in other sheet regions that do not intersect any filaments or massive halos. Our results are qualitatively similar using metalicity thresholds as large as $0.1\,Z_{\odot}$ and dark matter overdensity thresholds in the range $2.5-10$.

\smallskip
A detailed convergence study of the covering fraction of neutral hydrogen %gas clouds 
in the IGM in our five simulations with varying resolutions, accounting for different viewing angles through the sheet, will be presented in an upcoming study (Mandelker et al. in prep.). Here, we wish to highlight in \fig{convergence} the fact that metal-free LLSs in the IGM occur naturally in our simulations with sufficient resolution, with non-negligible covering fractions. Furthermore, it is interesting to note the decline in the covering fraction of dense clouds with redshift, by a factor of $\gsim 70$ from $z\sim 5$ when the clouds are formed following the sheet collision, to $z\sim 3$. This decline may be caused by cosmic expansion, which causes the pressure in the sheet to decline by a factor of $\gsim 5$ from $z=(5-3)$, in rough agreement with the naive scaling of $P\propto(1+z)^5$ in the IGM. As the cold clouds all have approximately the same temperature, the typical cloud density declines by a similar factor, causing the neutral fraction to decline by an even larger factor. Alternatively, cold clouds moving rapidly through a hot medium are expected to be disrupted on a %the so-called
cloud-crushing timescale, $t_{\rm cc}=2R_{\rm cl}v_{\rm cl}^{-1}\delta^{1/2}$, where $R_{\rm cl}$ is the cloud radius, $v_{\rm cl}$ its velocity, and $\delta$ is the density ratio between the cloud and the background \citep[e.g.][]{Agertz07}. In our case, $R_{\rm cl}\sim 2\kpc$, $v_{\rm cl}\sim 50\kms$ comparable to the sound speed in the post-shock medium, and $\delta\sim 30$ (\fig{phase_diagrams}). This yields $t_{\rm cc}\sim 0.2\Gyr$, while the time between $z=(5-3)$ is $\sim 1\Gyr$. This would imply that clouds are continuously created during this period, presumably due to turbulence in the %cooling
sheet. 
%These two options, and the issue of cloud formation and destruction more generally, will %be studied in detail in an upcoming paper (Mandelker et al, in prep.).

%%%%%%%%%%%%%%%%%%%%%%%%%%%%% 
\section{Discussion and Conclusions} 
\label{sec:conc}

\smallskip
We have presented a new cosmological simulation which zooms in on two massive, $\sim 5\times 10^{12}\, \msun$ halos connected by a Mpc-scale cosmic filament at $z\sim 2$. This large zoom-in region enables us to resolve the IGM, far from any massive halo, at unprecedented resolution. The simulation reveals the growth of the large scale structure and the cosmic web around our system. Two inclined cosmic sheets collide at $z\sim 5$ to produce a single massive sheet containing numerous dense filaments along which nearly all halos with $\Mv>10^9\msun$ lie. Following a major merger of one of our two main halos at $z\lsim 3$, the filaments merge leaving %behind the single massive filament we selected for our zoom-in region. 
a single massive filament. 

\smallskip
The sheet collision at $z\sim 5$ triggers a strong shock within the resulting sheet, heating the gas %up
to $T\sim 2\times 10^5\K$ with its pressure fixed by the ram pressure of infalling gas at $P\sim 50\K\cmc$. Due to thermal instability, this gas then separates into a volume-filling and a mass-dominating phase with $(n_{\rm H},T)\sim (10^{-4.5}\cmc,10^6\K)$ and $(10^{-3}\cmc,10^{4.5}\K)$ respectively.
%phase with $(n_{\rm H},T)\sim (10^{-4.5}\cmc,10^6\K)$, and a mass-dominating phase with $(n_{\rm H},T)\sim (10^{-3}\cmc,10^{4.5}\K)$. 
The cold phase is produced by shattering of the sheet %gas 
into cloudlets with sizes comparable to the local cooling length, $l_{\rm cool}=\cs t_{\rm cool}$ \citep{McCourt18}. In the post-shock gas, $l_{\rm cool}\sim 10\kpc$ and the typical cell size in %this region of 
our high resolution simulation is $800\pc$. We thus resolve the shattering of the sheet into kpc-scale fragments. While this is comparable to %the cooling length 
$l_{\rm cool}$ in the cold phase, we caution that the cloud sizes are likely influenced by our numerical resolution. These dense cloudlets result in high-column densities of neutral hydrogen %even 
in regions of the sheet which do not intersect any filaments or contain any halos with $\Mv>10^9\msun$. In particular, the covering fraction of pristine LLSs, with $N_{\rm HI}>10^{17.2}\cms$, 
$Z<10^{-4}\,Z_{\odot}$, and dark matter density less than 5 times the universal mean, is $\sim 1\%$ at $z\sim 3$ and 4 and $\sim 35\%$ at $z\sim 5$ when the sheet is viewed close to face-on. Whether individual cloudlets can be observationally disentangled from each other or from the filaments if the sheet were viewed close to edge-on, and whether such viewing angles might yield metal-free Damped Lyman-$\alpha$ Absorbers (DLAs, $N_{\rm HI}>10^{20.3}\cms$) will be presented in an upcoming paper (Mandelker et al. in prep.). Preliminary results suggest that the latter seems unlikely. 
%These LLSs have primordial composition, with $Z<10^{-4}\,Z_{\odot}$, and lie in regions where the dark matter density is less than 5 times the universal mean. 
%Owing to their large distance from any massive galaxy, this also

\smallskip
The large distance of these clouds from any massive galaxy implies that this result is likely robust to the adopted galaxy formation sub-grid physics. The production of metal-free LLSs, as recently observed by \citet{Fumagalli11a,Crighton16,Robert19}, thus seems to occur naturally in the IGM due to thermal instabilities induced by the growth of large-scale structure. This supports speculations that a growing fraction of LLSs at $z>3$ can be found in the IGM rather than the CGM around galaxies. 

\smallskip
In simulations with resolution comparable to the Illustris TNG100 simulation, the typical cell size in the sheet regions is $\sim 4\kpc$. Therefore, the cooling length in the post-shock gas is only marginally resolved, and %therefore 
the shattering process does not take place. As a result, the covering fraction of pristine LLSs is greatly reduced at $z\sim 5$ and is $\lsim 10^{-4}$ at $z\lsim 4$. This highlights the importance of achieving high-resolution in the IGM, even well outside the CGM of any galaxy. 

\smallskip
Another potential application of our results is with regard to the LyAF, and its relation to the underlying dark matter density distribution. The LyAF can be used to probe the mildly nonlinear matter power spectrum, down to scales of $<1$ comoving Mpc. At $z\sim 4-6$, where the physical scales probed are several tens of kpc, the power at these small scales has been used to rule out various warm dark matter (WDM) models \citep[e.g.][and references therein]{Viel13}, as a too light WDM particle would suppress power on these small scales compared to observations. These studies are often calibrated against cosmological simulations which have much lower resolution in the IGM than the simulations in this work. However, thermal instabilities such as those identified in this work may lead to excess power in the LyAF on scales of tens of kpc, corresponding to the cooling length of post-shock gas in cosmic sheets (\fig{phase_diagrams}), which is not associated with the underlying dark matter distribution. This may influence the constraints on WDM models from simulations which do not resolve the shattering. In addition to WDM constraints, our results may influence constraints on the temperature and the optical depth of the IGM at high redshift, which are also based on analysis of the LyAF \citep[e.g.][]{Viel13,Lidz14,Eilers18}.

%This topic will be explored in future work.

\smallskip
While the results presented in this paper are likely robust to the galaxy formation sub-grid physics, we caution against drawing too broad conclusions from the single system simulated in this study. In particular, it is currently unclear whether the shattering process discussed here requires a violent collision between sheets, or whether smooth accretion would have the same effect. It is also currently unknown how frequent such collisions between sheets are. Therefore, we cannot confidently extrapolate from the results presented here to the actual number density of metal-free LLSs in the IGM produced by shattering of cosmic sheets. Future simulations which employ similar methods of enhancing the resolution in the IGM will help to shed light on this problem.

%%%%%%%%%%%%%%%%%%%%%%%%%%%%%%%%% 
\section*{Acknowledgments} 
We thank the anonymous referee for constructive comments which helped improve the clarity of this manuscript. We thank Joe Hennawi, Siang Peng Oh, Drummond Fielding, Michele Fumagalli, and Ruediger Pakmor for helpful discussions. We acknowledge support from the Klauss Tschira Foundation through the HITS Yale Program in Astropysics (HYPA). F.C.v.d.B received additional support from the National Aeronautics and Space Administration through Grant No. 17-ATP17-0028 issued as part of the Astrophysics Theory Program.

%%%%%%%%%%%%%%%%%%%%%%%%%%%%%%%%%%%% 
\bibliographystyle{mn2e} 
%\bibliography{biblio}

\label{lastpage} 
 
\end{document}